\documentclass[a4paper,11pt]{article}
\pdfoutput=1 

\usepackage{jheppub} 

\usepackage{hyperref}
\usepackage{epsfig}
\usepackage{amssymb}
\usepackage{setspace}
\usepackage{amsmath}
\usepackage{amssymb}
\usepackage{graphicx}
\usepackage{subfigure}
\usepackage{url}
\usepackage{slashed}

\newcommand{\ie}{{\it i.e.}}

\newcommand{\cf}{{\it cf.\;}}

\newcommand{\be}{\begin{equation}}
\newcommand{\ee}{\end{equation}}
\newcommand{\br}{\begin{eqnarray}}
\newcommand{\bea}{\begin{eqnarray}}
\newcommand{\eea}{\end{eqnarray}}
\newcommand{\er}{\end{eqnarray}}
\newcommand{\ba}{\begin{array}}
\newcommand{\ea}{\end{array}}
\newcommand{\bi}{\begin{itemize}}
\newcommand{\ei}{\end{itemize}}
\newcommand{\bn}{\begin{enumerate}}
\newcommand{\en}{\end{enumerate}}
\newcommand{\bc}{\begin{center}}
\newcommand{\ec}{\end{center}}

\newcommand{\beq}{\begin{equation}}
\newcommand{\eeq}{\end{equation}}

\newcommand{\gsim}{\lower1.0ex\hbox{$\;\stackrel{\textstyle>}{\sim}\;$}}
\newcommand{\lsim}{\lower1.0ex\hbox{$\;\stackrel{\textstyle<}{\sim}\;$}}
\newcommand{\bs}{\begin{small}}
\newcommand{\es}{\end{small}}

\usepackage[T1]{fontenc} 

\title{Asking for an extra photon in Higgs production at the LHC and beyond}

\author[a]{Emidio Gabrielli,}
\author[b]{Barbara Mele,}
\author[c]{Fulvio Piccinini}
\author[d]{and Roberto Pittau}

\affiliation[a]{Dipart. di Fisica Teorica, Universit\`a di 
Trieste, Strada Costiera 11, I-34151 Trieste, Italy, \\ 
INFN, Sezione di Trieste, Via Valerio 2, I-34127 Trieste, Italy,\\
and NICPB, Ravala 10, Tallinn 10143, Estonia}
\affiliation[b]{INFN, Sezione di Roma, c/o Dipart. di Fisica, ``Sapienza" Universit\`a di Roma, \\ P.le Aldo Moro 2, I-00185 Rome, Italy}
\affiliation[c]{INFN, Sezione di Pavia, Via A. Bassi 6, I-27100 Pavia, Italy}
\affiliation[d]{ Departamento de F\'isica Te\'orica y del Cosmos and CAFPE, 
Universidad de Granada, \\ Campus Fuentenueva s.n., E-18071 Granada, Spain}

\emailAdd{emidio.gabrielli@cern.ch}
\emailAdd{barbara.mele@roma1.infn.it}
\emailAdd{fulvio.piccinini@pv.infn.it}
\emailAdd{pittau@ugr.es}

\abstract{We study  the inclusive production of a Higgs boson in association with a high-$p_T$ photon at the LHC, detailing the  leading-order features of the main processes contributing to the $H\gamma$ final state.   
Requiring an extra hard photon in Higgs production upsets the cross-section
hierarchy for  the dominant channels. The $H\gamma$ inclusive production comes mainly from photons radiated in vector-boson fusion (VBF), which accounts for about 2/3 of the total rate, for $p_T^{\gamma,j} >30$ GeV, at leading order. On the other hand, radiating a high-$p_T$ photon in the main  top-loop Higgs channel implies an extra parton  in the final state, which suppresses the production rate by a further $\alpha_S$ power. As a result, the 
$H\gamma$ production via top loops  at the LHC has  rates comparable with the ones arising from either the $H t\bar t$ production or  the $HW(Z)\gamma$ associated production. Then, 
in order of decreasing cross section,  comes the single-top-plus-Higgs channel, 
followed in turn by the 
 heavy-flavor fusion processes $b\bar b \to H\gamma$ and $c\bar c \to H\gamma$. The 
$H\gamma$ production  via
  electroweak loops  has just a minor role. \
  At larger c.m. energies, the $H t\bar t\gamma$ channel surpasses the total
  contribution of top-loop processes. 
  In particular, requiring $p_T^{\gamma,j} >30$ GeV at $\sqrt S \simeq 100$~TeV, $H t\bar t\gamma$  accounts for about $1/4$ of the inclusive $H\gamma$ production  at leading order, about half of the total  being due to VBF production.
}

\begin{document} 
\maketitle
\flushbottom

\section{Introduction}

The observation of a Higgs boson signal at the LHC \cite{Aad:2012tfa}  opened up a new era for collider physics. On the one hand, a major task of the LHC and future high-energy colliders is now to verify with high accuracy the actual properties of the new state, in order to check whether the standard model (SM) really provides the complete description of the electroweak symmetry breaking (EWSB) through the Higgs mechanism~\cite{Englert:1964et}
at the TeV energy scale, or some theory modification is needed. On the other hand, Higgs boson production in the SM can itself act as a background for new possible exotic states that might involve Higgs bosons in their production or decay channels. As a consequence, the most accurate predictions on both Higgs production mechanisms and Higgs decay characteristics in the SM are desirable.

In this paper, we discuss the Higgs production associated to a prompt high-$p_T$  photon at the LHC. 
 The $H\gamma$ final state can be experimentally quite distinctive,
 and might probe $H\gamma$ interactions in a nontrivial way.
After requiring an extra high-$p_T$ photon, the naive expectation is that  the original Higgs production mechanisms should be  suppressed by a few order of magnitudes, corresponding to an extra $\alpha$ factor in cross sections (where $\alpha=e^2/4\pi$),
while maintaining their relative weight.
Actually,  the main Higgs production mechanisms react in different ways to the requirement of photon radiation.
In particular, we will show here that the normal hierarchy  in the Higgs production channels  is upset by the requirement of an extra  high $p_T$ photon.
 
Higgs production at the LHC mainly proceeds, 
in order of decreasing rate, 
via gluon-gluon ($gg$) fusion (mostly through a top-quark loop), vector-boson fusion (VBF), associated 
$VH$ production (where $V$ is either a $W$ or a $Z$ boson), associated  $t\bar tH$ and $b\bar bH$ production, and single-top $tH$ production. Predictions for the corresponding cross sections have been worked out with good accuracy (including at 
least QCD NLO corrections for all processes \cite{H-xsect}). At the LHC, there is then a substantial  hierarchy in the corresponding cross sections, and the  gluon-fusion production via a top-loop turns out to be  by far the dominant contribution to the inclusive Higgs production, being an order of magnitude higher than the VBF
process,  as detailed in the following.

\begin{figure}
\begin{center}
\vskip -0.5cm
\hskip 0.4cm
\includegraphics[width=0.3\textwidth]{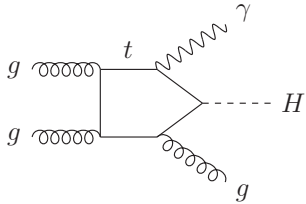}
\vskip -0.4cm
\caption{\small Basic top-quark pentagon diagram contributing to $gg\to H  \gamma \, g$.}
\label{fig_1}
\end{center}
\end{figure}

The request of an extra photon in the final state changes drastically the latter  ordering.
Indeed, the process $gg\to H\gamma$, occurring via a top-box diagram, is forbidden by Furry's theorem, and in general by $C$ parity. Then, the lowest-order partonic processes proceeding via QCD interactions (and unsuppressed by small Yukawa couplings)
are either the light-quark initiated processes $gq(\bar q)\to H  \gamma \, q(\bar q)$  and  
$q\bar q\to H  \gamma \, g$ (both involving a top-loop $ggH$ vertex)
or $gg\to H  \gamma \, g$ (via a pentagon top loop) [Figure~\ref{fig_1}].
The contribution of the $gg\to H  \gamma \, g$ amplitude to the $H \gamma j$
rate has been recently
evaluated in \cite{Agrawal:2014tqa}, where the latter process is claimed to be responsible for  the dominant
production of $H \gamma j$ final states at the LHC, followed  by the heavy-quark $Q\bar Q,gQ$ scattering into $H\gamma$.

In the present study,
 we will show that the $gg\to H  \gamma \, g$ channel actually contributes  
 to the inclusive $H \gamma j$ production to a much lesser extent 
 than previously stated. Indeed, on the one hand, the 
 $ggH$-loop-mediated production via 
 the $t$-channels $gq(\bar q)\to H  \gamma \, q(\bar q)$
  will be found to be about one order of magnitude larger than the one mediated by the top-pentagon amplitude $gg\to H  \gamma \, g$ at the LHC.
 On the other hand, the actual (by far) dominant production of $H \gamma $ final states accompanied by 
 jets will turn out to proceed through an electroweak process, that is the VBF Higgs production
 $q\bar q\to H  \gamma \, q\bar q$, where the high-$p_T$ photon radiation 
 by the initial/final quarks, connected by $W$ charged currents, is enhanced by the absence of suppressive QCD coherence effects \cite{Gabrielli:2007wf,Arnold:2010dx}.
 
 We will also evaluate for the first time the contributions to the inclusive $H\gamma$ production arising from  a hard photon radiated in the   
 associated  production    of a Higgs plus either a top-quark pair in 
 $Ht\bar t$  final states or a single top in the  $t$-channel  $Ht\,(H\bar t)$ production. The $H t\bar t\gamma$ will be found to 
 contribute at the LHC 
 at the same level as the  $gq(\bar q)\to H  \gamma \, q(\bar q)$ channels. Remarkably, the relative $H t\bar t\gamma$ weight increases at larger c.m. energies, approaching the relative (still dominant) contribution of the  VBF component
 at $\sqrt S\sim100$ TeV.
 The $H t\gamma$ component is of course lower than the $H t\bar t\gamma$ one, but 
 less  than naively expected, thanks to the enhancing mechanism related to the  
 $W$-exchange in the $t$-channel $H t$ production, similar to the one acting in the  VBF case \cite{Gabrielli:2007wf,Arnold:2010dx}.
 
 The $HV\gamma$ (with $V=W,Z$) associated production has been evaluated in \cite{Mao:2013dxa,Shou-Jian:2015sta} at the NLO in QCD. As we will see, at the LHC energies, it also contributes to the $H \gamma$ production in a comparable way to  the $t$-channels $gq(\bar q)\to H  \gamma \, q(\bar q)$ (and to the $H t\bar t\gamma$ process).
 
 We then reconsider the total contribution to the $H\gamma$ final states of the heavy-flavour ($b\bar b$ and $c\bar c$) scattering~\cite{Abbasabadi:1997zr}, that will be found  comparable to the $gg\to H  \gamma \, g$
 process at the LHC.
 We finally comment on the minor $H\gamma$ component due  the
 $q \bar q \to H\gamma$ electroweak-loop process previously studied in \cite{Abbasabadi:1997zr}.

 We stress that the present study is not aimed to provide the most accurate estimate of the production cross sections for different $H\gamma$ channels, but rather to analyze the inclusive production of the $H\gamma$ system through its main components, discussing the corresponding relative weight. Such a discussion has non-trivial aspects that, to our knowledge, have not previously been detailed in the literature.

 The plan of the paper is the following. In Section 2,
 we discuss the cross-section computation for the  different channels 
 contributing to the  $H\gamma$ final state in proton-proton collisions. 
For some processes, a QCD next-to-leading-order (NLO) evaluation is available,
for others,  even the tree-level estimates are, to our knowledge, still missing
in the literature, and we provide them here.  In particular, we present LO $H\gamma$ cross sections relevant for LHC and future higher-energy $pp$ colliders.
 In Section 3, we compare the different contributions, looking at both total rates and 
 Higgs/photon kinematical distributions. Hierarchies of $H\gamma$ cross sections are then  compared
 with the ordering of original Higgs production mechanisms.  In Section 4, we present our conclusions 
 and outlook. In the Appendix, we report the 
 asymptotic behavior of the top pentagon amplitude for the $gg\to H \gamma g$ channel.

 \section{Processes contributing to the $H\gamma$ final state}
 
 In this section, we detail the present theoretical knowledge of the various channels contributing to the associated production of a Higgs boson and a high-$p_T$ photon at the LHC. For a few of them QCD NLO predictions are available, others have been computed only at leading order (LO), while some processes like  the Higgs production in association with top quarks to our knowledge have not  yet been considered in the literature. We will discuss the main processes in order of decreasing relevance of the corresponding channels with no photon emission, which are  responsible for the dominant Higgs boson production at the LHC.
Remarkably, we will see 
 that the request of an extra high-$p_T$ photon in the basic Higgs production processes will have a strong impact on the relative weight  of different channels.
 
When quoting the $H\gamma$ production rates in the present study, we will assume  
common sets of input parameters and  kinematical cuts. The latter are needed in most of the channels considered, which are characterized by collinear- and soft-photon 
(and -parton) divergencies. The setup applied in all cross-section computations 
(even in absence of divergencies) is 
 \bea
p_T^\gamma&>&30 {\;\rm GeV}, \;\;\;\;\;\;\; |\eta_\gamma| < 2.5\, ,  \nonumber \\
p_T^j&>&30 {\;\rm GeV},\;\;\;\;\;\;\; |\eta_j| < 5 \, , \; \nonumber\\
\Delta R(\gamma,j_i)&>&0.4,\;\;\;\;\;\;\; \Delta R(j_1,j_2)>0.4\, , 
\label{cuts}
\eea
where $\Delta R(a,b)=\sqrt{\Delta \phi(a,b)^2+\Delta \eta(a,b)^2}$
is the angular separation between $a$ and $b$, and $j_i \,(i=1,2)$ is any parton in the final state.
We then set the Higgs and the heavy quark  masses as follows :
\bea
m_H&=&125 {\;\rm GeV}, \;\;\;\;\;\;\; m_t= 173  {\;\rm GeV}, \; \nonumber \\
m_b^{\overline{MS}}(m_H)&=& 2.765 {\;\rm GeV}, \;\;\; m_c^{\overline{MS}}(m_H)=0.616 {\;\rm GeV},
\label{masses}
\eea
where we assumed the running masses at the $m_H$ scale in the Yukawa couplings
entering the $b\bar b, c \bar c \to H\gamma$ cross sections.

In the present analysis, we are mainly interested in establishing the relative importance of the main processes giving rise to $H\gamma$ final states. Since,  
for most of the channels we will analyze in the following, QCD NLO cross sections have not yet been computed, in order to make a consistent comparison, 
  we will always consider QCD LO rates (even when  QCD NLO  estimates are  already available in the literature).
  
  When considering the absolute rates that we will present 
below,  one should then keep in mind the influence of the  cut choice on the cross sections. In Section 3,
we will show how   relaxed  requirements on kinematics can influence the relative weight of different channels.

  To compute LO cross sections and distributions we use {\tt AlpGen}~\cite{Mangano:2002ea}\footnote{Since not all processes treated 
in this study are available in the official release {\tt v2.14}, we have extended the code, to include 
also the processes $gg, q {\bar q} \to H \gamma t \bar t$, $b {\bar q} \to H \gamma t {\bar q}$, 
$Q {\bar Q} \to H \gamma$ and $Q {\bar Q} \to H$ with $Q=b,c$. }, 
with the parton distribution set CTEQ5L \cite{Lai:1999wy}, setting the  factorization $(\mu_F)$ and renormalization $(\mu_R)$ 
scale at the common value $\mu=m_H$.
We will estimate the LO cross-section uncertainties by varying $\mu=\mu_F=\mu_R
$ in the range $[\frac{1}{2} \, m_H,2\, m_H]$. 

In the next subsections, we will report
cross sections for the different channels according  to the parameters and settings
described above, for proton collision  c.m. energies of 14 TeV, 33 TeV and 100 TeV,
the latter being relevant for Future Circular Collider (FCC) studies that are presently under 
way~\cite{fcc}\footnote{Physics and, in particular, Higgs physics at 100 TeV 
have been recently reviewed in \cite{Arkani-Hamed:2015vfh} and \cite{Baglio:2015wcg}, respectively.}. 
Proton collisions at 33 TeV could  be realized in an upgraded-energy program of the LHC (usually named HE-LHC \cite{Andreazza:2015bja}).
%
%
%
%
%

%
 \subsection{QCD production via top loops in $gg, qg (\bar q g),\bar q q \to H\gamma\, j$}	
\begin{figure}
\begin{center}
\vskip -0.6cm
\hskip 0.1cm
\includegraphics[width=0.4\textwidth]{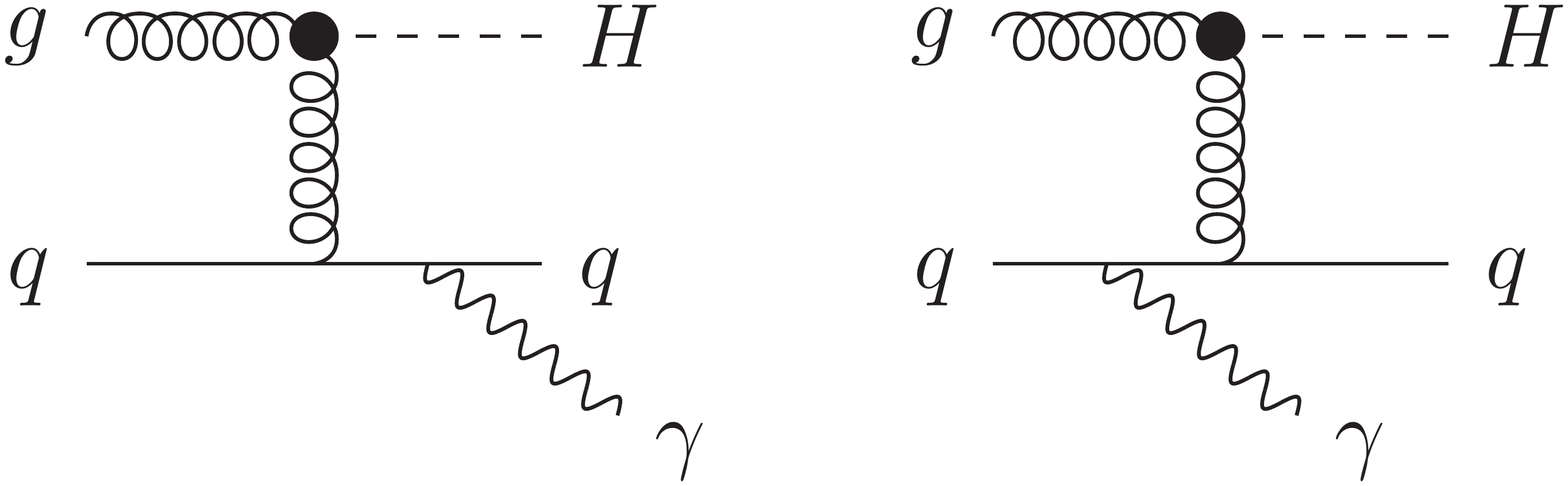}
\vskip -1.5cm
\caption{\small Feynman diagrams for $gq \to H \gamma q $.
The black blob represents the $ggH$ effective vertex.}
\label{fig_2}
\end{center}
\end{figure}

\begin{figure}
\begin{center}
\vskip -0.8cm
\hskip 0.1cm
\includegraphics[width=0.4\textwidth]{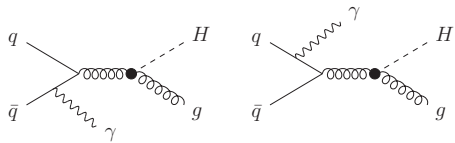}
\vskip -1.5cm
\caption{\small Feynman diagrams for $q\bar q \to H \gamma g $. 
The black blob represents the $ggH$ effective vertex.}
\label{fig_3}
\end{center}
\end{figure}

 When asking for an extra photon in the main Higgs production channel $gg\to H$ proceeding  via gluon fusion into a top triangle loop, one is forced to pass to the next QCD order, and include an extra parton in the collision final state. Indeed, 
 as already mentioned, Furry's theorem forbids the emission of a photon from the $gg H$ top-quark loop, and the $gg\to H\gamma$ amplitude vanishes\footnote{Note that the vanishing of the $gg\to H\gamma$ amplitude is a general consequence of $C$ parity,
 which holds in any SM theory extension.}.  
 Then, one can either require a further gluon emission in the latter process, and obtain a non-vanishing $gg\to H\gamma g$ amplitude via a top pentagon loop 
 (Figure~\ref{fig_1}), or ask for an extra photon radiation in the Higgs+jet production   proceeding via   the channels 
 $gq \to H q $ 
 $(g\bar q  \to H\bar q)$, and 
 $\bar q q \to Hg$ by means of 
 a top triangle loop (Figure~\ref{fig_2} and \ref{fig_3}, respectively). 

 The production of 
 $H\gamma j $ final states from the gluon fusion $gg\to H\gamma g$ channel at hadron colliders has been recently studied in \cite{Agrawal:2014tqa}. The $gg\to H\gamma g$  amplitude is  gauge invariant and finite, and 
 one can compute the  gluon-fusion separate contribution to $H\gamma$
 production. In our analysis, we have redone the evaluation of the top pentagon
 amplitude ${\cal A_{\rm Pent}}$ associated to the  $gg\to H \gamma g$ channel. 
  We detail our computation in the following.
 
 ${\cal A_{\rm Pent}}$ is given by the sum of the 24 pentagon-like diagrams obtained by permuting in all possible ways the external vectors in Figure~\ref{fig_1}. Each diagram can be expressed in terms of a linear combination of one-loop scalar boxes, triangles, bubbles, massive tadpoles and rational terms.  The coefficients of all  scalar functions have been obtained numerically 
via the OPP approach~\cite{Ossola:2006us}, as implemented in {\tt CutTools}~\cite{Ossola:2007ax}, linked to the one-loop scalar functions in~\cite{vanHameren:2010cp}.
As each pentagon is separately ultraviolet (UV) convergent, no rational term of the $R_2$  kind is present~\cite{Ossola:2008xq}. Thus, the full rational part of ${\cal A_{\rm Pent}}$ is $R_1$-like, and also numerically provided by {\tt CutTools}.
 
The input needed by {\tt CutTools} is the integrand of each diagram as a function of the integration momentum. In order to speed up the computation, we have used an in-house implementation of the massive helicity method~\cite{Kleiss:1985yh} that expresses traces over gamma matrices in terms of scalar products in the spinor space. This gives a numerical stable answer for most of the phase-space points. In order to detect and rescue the remaining unstable configurations, we have used the built-in quadruple-precision facilities  of {\tt CutTools}. As a result, no randomly generated phase-space point is discarded during the Monte Carlo integration. As for the latter, the value of $|{\cal A_{\rm Pent}}|^2$ computed by {\tt CutTools} is plugged into a code based on {\tt AlpGen}, which, besides integrating over the relevant phase-space, also takes care of the convolution with the gluon parton densities\footnote{The corresponding numerical code is available in 
 http://www.ugr.es/$\sim$pittau/PENTAGON/.}. 

In order to validate the correctness of our calculation, a numerical check of gauge invariance has been performed. In particular, 
 we have numerically replaced polarization vectors by four-momenta, obtaining zero up to the machine precision. In addition, we have checked the numerical agreement between the result obtained when using a large input value for the top mass and the analytic asymptotic behavior of ${\cal A_{\rm Pent}}$, as reported in the Appendix. 

Finally, we have compared our outcome for the $gg\to H\gamma g$ channel with the results  in~\cite{Agrawal:2014tqa}, and 
 found complete agreement on both numerical cross sections and kinematical distributions. We stress that there is no infrared divergence for either photons or gluons in the final state, and the maximum of the corresponding $p_T$ distributions  is ruled by the top mass circulating in the pentagon loop.
 In particular, the photon and gluon $p_T$ distributions are both peaked at $p^{max}_T\!\!\sim 120 $ GeV at 14 TeV, while the Higgs  distribution is maximal at $p_T^{max}\!\! \sim 80 $
 GeV~\cite{Agrawal:2014tqa}.
 
 The $gg\to H\gamma g$ cross section corresponding to the setup in Eqs.~(\ref{cuts})-(\ref{masses})  is
\bea
\sigma(gg\to H\gamma g)^{\sqrt S=14 \, {\rm TeV}}&=& 0.287^{\;+0.138}_{\;-0.086}  \;{\rm fb} \label{ggh14} \, , \\
\sigma(gg\to H\gamma g)^{\sqrt S=33 \, {\rm TeV}}&=& 1.79^{\;+0.71}_{\;-0.47}  \;{\rm fb}  \, ,\label{ggh33} \\
\sigma(gg\to H\gamma g)^{\sqrt S=100\,  {\rm TeV}} &=& 12.0^{\;+3.6}_{\;-2.6} \; {\rm fb} \, ,
\label{ggh100}
\eea
where the cross section {\it central value} assumes  $\mu_F=\mu_R=m_H$, and the upper (lower) variations correspond to $\mu_F=\mu_R=\frac{1}{2} m_H$ 
($\mu_F=\mu_R=2 \,m_H$).

%
%

%
%

 In~\cite{Agrawal:2014tqa}, the $gg\to H\gamma g$  rate has been compared to the
 heavy-quark $Q\bar Q+Qg \to H\gamma j$ 
 cross section (with $Q=b,c$),  
 and claimed to provide the dominant contribution to the 
 $H\gamma j$ final state at hadron colliders.
 Here, we correct the latter statement, by including also the $H\gamma j $ production
 initiated by light quarks, proceeding via the top triangle  $ggH$ vertex in 
 either  the $t$-channel (Figure~\ref{fig_2}) or  the $s$-channel (Figure~\ref{fig_3}). 
 The corresponding LO cross section (summing up over the $t$ and $s$ channels, and assuming the same set of cuts  and conventions as above) have been obtained by 
  {\tt AlpGen},  by a $ggH$ effective vertex :
 \bea
\sigma(gq,g\bar q,q\bar q\to H\gamma \, q,\bar q,g
)^{\sqrt S=14 \, {\rm TeV}}&=& 2.77^{\;+0.40}_{\;-0.34}  \;{\rm fb}  \, ,\label{gqh14}\\
\sigma(gq,g\bar q,q\bar q\to H\gamma \, q,\bar q,g
)^{\sqrt S=33 \, {\rm TeV}}&=& 11.1^{\;+1.1}_{\;-0.9}  \;{\rm fb}  \, ,\label{gqh33}\\
\sigma(gq,g\bar q,q\bar q\to H\gamma \, q,\bar q,g)^{\sqrt S=100\,  {\rm TeV}} &=& 54.0^{\;+1.9}_{\;-2.0} \; {\rm fb}\, .
\label{gqh100}
\eea
Note that the $s$-channel $q\bar q\to H\gamma g$ cross section provides a tiny component to the latter rates, amounting to about 2.8\% of the total cross section at 14 TeV, and 1.9\% at 100 TeV.

As a result, at the 14-TeV LHC, the light-quark initiated contribution to the $H\gamma j $ production
turns out to be  an order of magnitude larger than the pentagon gluon-fusion production
(\cf~Eqs.~(\ref{ggh14}) and (\ref{gqh14})).

%

\subsection{Vector boson fusion}
\begin{figure}
\begin{center}
\vskip -0.6cm
\hskip 0.1cm
\includegraphics[width=0.4\textwidth]{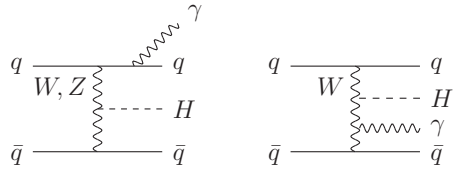}
\vskip -1.cm
\caption{\small Two representative diagrams for the VBF channel 
$q\bar q\to H  \gamma \, q\bar q$.}
\label{fig_4}
\end{center}
\end{figure}
The Higgs-photon associated production in VBF 
is obtained by the emission of a high-$p_T$ photon  from  either the initial/final quarks or  the $t$-channel $W$ propagators, as shown in Figure~\ref{fig_4} for two representative diagrams out of the complete set.
The  $q\bar q\to H  \gamma \, q\bar q$ channel\footnote{The possibility of different quark flavors in  initial and final states is understood in our notation.}
 has been studied at LO in \cite{Gabrielli:2007wf}, and   at NLO 
 in \cite{Arnold:2010dx}. 
Apart from the setup detailed in Eqs.~(\ref{cuts})-(\ref{masses}), we will assume a further cut on the quark-pair invariant mass, $M_{j_1,j_2}>100$~GeV, hence depleting the contribution from the $q\bar q\to H W/HZ$ associated production, which will be  considered separately  in the following. 

Asking for an extra high-$p_T$ photon in VBF drastically increases the relative importance of the $WW$ fusion component with respect to the $ZZ$ one~\cite{Gabrielli:2007wf}.  Indeed,  in the $ZZ$ fusion channel,  destructive-interference effects occur  between 
the photon radiation from initial and final quarks connected by a $t$-channel $Z$ exchange. As a result, while asking for an extra central photon with $p_T\gsim 20$~GeV 
typically suppresses the  $WW$-fusion cross section by two orders of magnitude, the corresponding decrease in the $ZZ$ component  is ${\cal O} (10^{-3})$. This makes 
the $ZZ$ fusion contribution to the total $q\bar q\to H  \gamma \, q\bar q$ cross section even smaller than naively expected, and almost negligible~\cite{Gabrielli:2007wf}. 

The total LO cross section for $q\bar q\to H  \gamma \, q\bar q$, computed by 
  {\tt AlpGen}, is
  \bea
\sigma( q \bar q\to H\gamma \, q \bar q)^{\sqrt S=14 \, {\rm TeV}}&=& 22.0^{\;+1.3}_{\;-1.1}  \;{\rm fb} \, , \label{VBF14} \\
\sigma( q \bar q\to H\gamma \, q \bar q)^{\sqrt S=33 \, {\rm TeV}}&=& 87.3^{\;+0.3}_{\;-0.0}  \;{\rm fb}  \, ,\label{VBF33} \\
\sigma(q \bar q\to H\gamma \, q \bar q)^{\sqrt S=100\,  {\rm TeV}} &=& 325.^{\;-23}_{\;+20} \; {\rm fb}\, .
\label{VBF100}
\eea
%

The VBF contribution is then found to be by far dominant over the top-loop mediated 
channels contributing to $H\gamma$ final states. In particular, $\sigma( q \bar q\to H\gamma \, q \bar q)$ is almost an order of magnitute larger than $\sigma(gq,g\bar q,q\bar q\to H\gamma \, q,\bar q,g
)$ at the LHC (\cf~Eq.~(\ref{gqh14})), and six times higher at 100 TeV 
(\cf~Eq.~(\ref{gqh100})).
Note that here we are applying a $p_T>30$GeV requirement on forward jets, that is quite stricter than the 20-GeV cut  usually applied in VBF studies at the LHC, hence considerably reducing the predicted cross sections. In Section 3, we will discuss
the effect of relaxing the relevant cuts on transverse momenta.

Contributions to the $H\gamma j j$ production different from VBF  can arise from  the NLO treatment of the $\,gg, qg (\bar q g),\bar q q \to H\gamma\, j$ channels 
analyzed in Section 2.1. These are expected 
to be quite depleted with respect to VBF~\cite{Gabrielli:2007wf}, and will not be considered
in this analysis.

\subsection{Associated $HW$ and $HZ$ production}
\begin{figure}
\begin{center}
\vskip -0.6cm
\hskip 0.1cm
\includegraphics[width=0.5\textwidth]{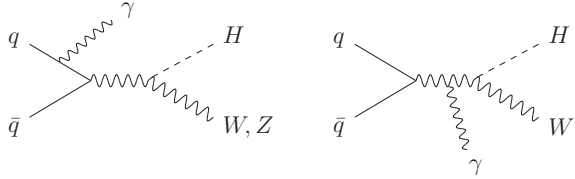}
\vskip -2cm
\caption{\small Two representative diagrams for the $HV\gamma$ associated production.}
\label{fig_5}
\end{center}
\end{figure}
The $H\gamma$ final states arising from the associated production
$q \bar q \to H W$, $q \bar q \to H Z$ derive from the emission of a hard photon
from the $q \bar q$ initial state, and, in the $HW$ case, from either the $W$ propagator 
or the final $W$ (see Figure~\ref{fig_5} for two representative diagrams out of the complete set). NLO predictions for $q \bar q \to H \gamma W$ and $q \bar q \to H \gamma Z$ have been presented in \cite{Mao:2013dxa} and \cite{Shou-Jian:2015sta}, respectively (see also \cite{Torrielli:2014rqa}).

Our {\tt AlpGen}  estimates for the corresponding LO cross sections are 
 \bea
\sigma( q \bar q\to H\gamma W)^{\sqrt S=14 \, {\rm TeV}}&=& 1.87^{\;-0.03}_{\;+0.02}  \;{\rm fb}  \, ,\label{HW14} \\
\sigma( q \bar q\to H\gamma W)^{\sqrt S=33 \, {\rm TeV}}&=& 5.19^{\;-0.37}_{\;+0.32}  \;{\rm fb}  \, ,\label{HW33} \\
\sigma(q \bar q\to H\gamma W)^{\sqrt S=100\,  {\rm TeV}} &=& 16.5^{\;-2.2}_{\;+2.1} \; {\rm fb}\, ,
\label{HW100}
\eea
and
 \bea
\sigma( q \bar q\to H\gamma Z)^{\sqrt S=14 \, {\rm TeV}}&=& 1.34^{\;-0.03}_{\;+0.03}  \;{\rm fb}  \, ,\label{HZ14} \\
\sigma( q \bar q\to H\gamma Z)^{\sqrt S=33 \, {\rm TeV}}&=& 3.49^{\;-0.28}_{\;+0.23}  \;{\rm fb}  \, ,\label{HZ33} \\            
\sigma(q \bar q\to H\gamma Z
)^{\sqrt S=100\,  {\rm TeV}} &=& 10.3^{\;-1.4}_{\;+1.4} \; {\rm fb}\, .
\label{HZ100}
\eea
At the LHC, the sum of the latter contributions amounts roughly to the total top-loop induced cross sections in Eqs.~(\ref{ggh14}) and (\ref{gqh14}).

%
%
%
%

\subsection{Top-pair and single-top final states}
\begin{figure}
\begin{center}
\vskip -0.6cm
\hskip 0.1cm
\includegraphics[width=0.5\textwidth]{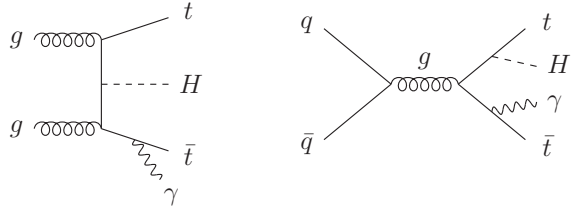}
\vskip -2cm
\caption{\small Two representative diagrams for $gg, q \bar q\to H\gamma \,t\bar t$.}
\label{fig_6}
\end{center}
\end{figure}
 The Higgs production via top-pair final states offers the unique opportunity to directly test the top Yukawa coupling. This channel is quite depleted  with respect to the $gg$, VBF, and $HV$-associated 
 production because of  $m_t$ phase-space effects. Requiring an extra hard photon 
 in the $Ht\bar t$ final states affects this hierarchy, since high-$p_T$ photons are more naturally radiated in the production of (more spherical) massive charged systems
 (see Figure~\ref{fig_6} for two representative diagrams out of the complete set).
 The $H\gamma \,t\bar t$  cross section computed at LO via {\tt AlpGen} is 
 \bea
\sigma(gg, q \bar q\to H\gamma \,t\bar t)^{\sqrt S=14 \, {\rm TeV}}&=& 2.55^{\;+0.89}_{\;-0.60}  \;{\rm fb} \, ,\label{Htt14} \\
\sigma(gg, q \bar q\to H\gamma \,t\bar t)^{\sqrt S=33 \, {\rm TeV}}&=& 17.8^{\;+5.4}_{\;-3.8}  \;{\rm fb} \, ,\label{Htt33} \\     
\sigma(gg, q \bar q\to H\gamma \,t\bar t
)^{\sqrt S=100\,  {\rm TeV}} &=& 159.^{\;+37}_{\;-29} \; {\rm fb}\, .
\label{Htt100}
\eea
%
%
%

The LHC $H\gamma \,t\bar t$ cross section turns out to be in the same ballpark of the top-loop $H\gamma j$ cross section, and also of the total $H\gamma V$ cross section (with $V=W,Z$). At larger $\sqrt S$, the $H\gamma \,t\bar t$ rate gets the upper hand, and approaches 
the VBF $H\gamma \, q \bar q$ rate. At 100 TeV, 
the $H\gamma \,t\bar t$ rate  is just about half the VBF $H\gamma \, q \bar q$ one,
 and we will see that the $H\gamma \,t\bar t$ channel becomes the second most important production 
 mechanism for $H\gamma$ final states.

The Higgs production associated to a single top is an electroweak process that can proceed via three
different channels at hadron colliders \cite{Maltoni:2001hu}. Here, we restrict to  the $t$-channel $b\bar q\to t H \bar q$  which has the largest cross section. The $t$-channel rate is anyway quite small
at the LHC. Nevertheless, its role has recently been emphasized for its sensitivity to a possible change in the relative sign of the $ttH$ and $WWH$ couplings \cite{Biswas:2012bd}.

\begin{figure}
\begin{center}
\vskip -0.6cm
\hskip 0.1cm
\includegraphics[width=0.5\textwidth]{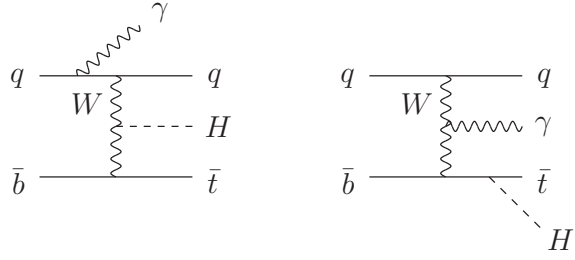}
\vskip -1.5cm
\caption{\small Two representative diagrams for $\bar b \, q\to H\gamma \,\bar t \,q$.}
\label{fig_7}
\end{center}
\end{figure}

In  Figure~\ref{fig_7}, one can find two representative diagrams out of the complete set for 
$b\bar q\to H \!\gamma t \bar q$. The $W$ exchange in the $t$-channel gives rise to a radiative pattern similar to the one in $WW$ fusion, where photon radiation from different quark legs does not interfere destructively. On the other hand,  the photon radiation 
somehow weakens the original cancellation among the different components of the 
$b\bar q\to t H \bar q$ amplitude~\cite{Maltoni:2001hu}. 

As a consequence, 
the requirement of an  extra $p_T>30$GeV  photon in the $b\bar q\to t H \bar q$ channel makes the cross section drop only 
by an amount ${\cal O} (10^{-2})$.
In particular,  
the {\tt AlpGen} estimate for the LO $\,b\bar q\to t H \bar q$ cross section is 
\bea
\sigma(b \, q\to H\gamma \,t \,q
)^{\sqrt S=14 \, {\rm TeV}}&=& 0.537^{\;-0.030}_{\;+0.016}  \;{\rm fb} \, , \label{Ht14} \\
\sigma(b \, q\to H\gamma \,t \,q
)^{\sqrt S=33 \, {\rm TeV}}&=& 4.19^{\;-0.42}_{\;+0.28}  \;{\rm fb} \, , \label{Ht33} \\     
\sigma(b \, q\to H\gamma \,t \,q
)^{\sqrt S=100\,  {\rm TeV}} &=& 29.8^{\;-4.5}_{\;+3.8} \; {\rm fb}\, ,
\label{Ht100}
\eea
where  $b  q\to H\!\gamma t q$ stands for a sum over the two charge-conjugated channels. 

%

%
%

\subsection{Heavy-quark $b\bar b,c\bar c$ fusion}
\begin{figure}
\begin{center}
\vskip -0.6cm
\hskip 0.1cm
\includegraphics[width=0.5\textwidth]{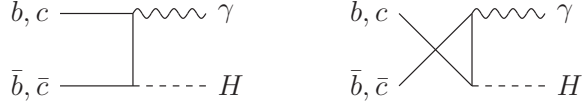}
\vskip -2.5cm
\caption{\small Feynman diagrams for $b\bar b, c\bar c \to H \gamma  $.}
\label{fig_8}
\end{center}
\end{figure}

Heavy-flavor quark annihilation, where initial $b\bar b$ or $ c\bar c$ pairs come
from the sea parton distributions, is 
the  lowest-order channel  producing a Higgs boson at hadron colliders. Despite the $b$- and $c$-quark Yukawa-coupling suppression,
after requiring an extra photon in the $b\bar b, c\bar c\to H$ channels (Figure~\ref{fig_8}), one still gets interesting cross sections. The one-order-of-magnitude difference in the 
$b\bar b\to H$ and  $ c\bar c\to H$ cross sections at the LHC 
(where the coupling ratio $[m_b^{\overline{MS}}(m_H)/m_c^{\overline{MS}}(m_H)]^2\sim 20$ (\cf  Eq.~(\ref{masses})) is partly compensated by the larger
$c$-parton distribution)  is reduced by a  factor 
$(Q_b/Q_c)^2=1/4$ in the  $b\bar b, c\bar c\to H\gamma $ cross sections, as will be shown in  Section 3.
The $b\bar b \to H\gamma$ cross section has been evaluated in the SM in  \cite{Abbasabadi:1997zr}, and in supersymmetric extensions of the SM in \cite{Gabrielli:2007zp}.

Our  {\tt AlpGen} estimate in the five-flavor scheme (5FS),  with running $b$ and $c$ masses evaluated  at the
$m_H$ scale in the Yukawa couplings, gives as a result
 \bea
\sigma(b \, \bar b\to H\gamma 
)^{\sqrt S=14 \, {\rm TeV}}&=& 0.162^{\;-0.041}_{\;+0.040}  \;{\rm fb}  \, ,\label{bb14} \\
\sigma(b \, \bar b\to H\gamma 
)^{\sqrt S=33 \, {\rm TeV}}&=& 0.713^{\;-0.202}_{\;+0.206}  \;{\rm fb}  \, ,\label{bb33} \\
\sigma(b \, \bar b\to H\gamma
)^{\sqrt S=100\,  {\rm TeV}} &=& 3.51^{\;-1.10}_{\;+1.20} \; {\rm fb}\, ,
\label{bb100}
\eea
and
%
 \bea
\sigma(c \, \bar c\to H\gamma 
)^{\sqrt S=14 \, {\rm TeV}}&=& 0.072^{\;-0.011}_{\;+0.010}  \;{\rm fb}  \, ,\label{cc14} \\
\sigma(c \, \bar c\to H\gamma 
)^{\sqrt S=33 \, {\rm TeV}}&=& 0.287_{\;+0.052}^{\;-0.053}  \;{\rm fb}  \, ,\label{cc33} \\
\sigma(c \, \bar c\to H\gamma
)^{\sqrt S=100\,  {\rm TeV}} &=& 1.28^{\;-0.29}_{\;+0.30} \; {\rm fb}\, .
\label{cc100}
\eea
Note that, although the $gg, q \bar q\to H \,t\bar t$ and $b\bar b\to H$ cross sections are
 comparable at 14 TeV (\cf Table \ref{tab:total}, next Section), the request of an extra photon depletes 
the $b\bar b\to H$ with respect to  not only the $gg, q \bar q\to H \,t\bar t$ channel,  but also the single-top $b q\to t H q$ process  (\cf  Eqs.~(\ref{Htt14})
and (\ref{Ht14})). 
%
%
%
\subsection{Electroweak  $\bar q q\to H\gamma$ production}
\begin{figure}
\begin{center}
\vskip -0.6cm
\hskip 0.1cm
\includegraphics[width=0.3\textwidth]{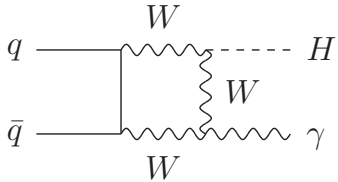}
\vskip -0.5cm
\caption{\small One representative  diagram for $\bar q q\to H\gamma$.}
\label{fig_9}
\end{center}
\end{figure}
A further channel mildly contributing to the $H\gamma$ associated production is
$q\bar q \to H\gamma$ that occurs via light-quark annihilation, going through
$s$-channel $\gamma$ and $Z$ exchange, involving a  $\gamma\gamma H$
and $Z\gamma H$ triangle loop of top quarks or $W$'s, and box diagrams with $W$'s and light quarks running in the loop.
Figure~\ref{fig_9} shows one diagram out of the complete set.
The corresponding cross sections have been computed at the Tevatron and the LHC
 \cite{Abbasabadi:1997zr}, and found  to be  quite smaller than the heavy-flavor tree-level $Q\bar Q \to H\gamma$ contribution to the
$H\gamma$ final state at the LHC.
We will then neglect the corresponding rates in the present discussion.

\section{Comparison of rates and distributions}
\begin{table}
\begin{center}
\begin{tabular}{l|c|c|c|c|c|c|}
$\sigma_{(p_T^{\gamma,j} >30 {\rm GeV})}$
 & ($H)_{\rm14TeV}$ &  ($H\gamma)_{\rm14TeV}$ 
 & ($H)_{\rm33TeV}$ &  ($H\gamma)_{\rm33TeV}$
 & ($H)_{\rm100TeV}$&   ($H\gamma)_{\rm100TeV}$ \\ \hline

 $gg,gq,q \bar q $ & \;\;30.8 pb & \;\;\;\; 3.05 fb & 137. pb &\,  12.9 fb 
 & \; 745.  pb & \;\;\; 65.8 fb\\ 
 
 VBF & 2.37 &  22.0 \;\;      & 8.64 &  87.3 \;\;
 & \;\, 31.0 \;\;& 325. \;\;\\
 
 $WH$ & 1.17 &  1.88       & 3.39 &  5.20
 & 12.1 & 16.6 \\

$ZH$ & 0.625 & 1.35          & 1.82 & 3.49
& \; \;6.52 & 10.3 \\

$t\bar tH$ & 0.585 & 2.55       &  4.08  &  17.8  \;\;
& \; 34.3\;\; & 158.\;\;\; \\ 

$tH+\bar tH$ &  0.056 & 0.536        &  0.428  &   4.17
&\;\;  2.18 & 29.7 \\ 

$b\bar b\to H$ & 0.670 & 0.162          &  2.82  &   0.713
& \;\, 14.6 \,\; & \;\;\; 3.51 \\ 

$c\bar c\to H$ & 0.069 & 0.072          &   0.265  &   0.287
& \;\; 1.20 & \;\;\; 1.28 \\  \hline
\end{tabular}
\end{center}
\caption{\small Cross sections  for $H\gamma$ associated production in $pp$ collisions 
at 14 TeV, 33 TeV, and 100 TeV, for the dominant channels, for $p_T^{\gamma,j} >30$ GeV. For comparison, also the cross sections for the corresponding channels without hard-photon radiation are reported, 
where the first row ($gg,gq,q \bar q $) refers to the sum of the Higgs and
Higgs-plus-one-jet contributions (see text). All cross sections are at LO, and computed via {\tt AlpGen}.  The complete set of selection cuts applied is described in the text.
}
\label{tab:total}
\end{table}

%
%
%
%
%
%
%
%
%
%
%

In the previous section, we detailed  the LO cross sections for the dominant $H\gamma$ production channels at different c.m. energies. We included a study of the scale dependence in order to get a flavor of 
  NLO correction effects. We are now going to discuss how the LO central values (\ie,  computed for $\mu=m_H$) for cross sections  of different processes compare, in order to pinpoint the main components
of the $H\gamma$ inclusive production. We also confront the cross sections of various $H\gamma$ channels with the cross sections of the corresponding Higgs production channels where no high-$p_T$ photon is radiated.
This will make manifest the fact that the presence of an extra photon in the final state deeply affects the hierarchy of importance for  Higgs production channels.

 In Table \ref{tab:total}, we show, for  $\sqrt S= $14 TeV, 33 TeV, and 100 TeV, LO cross sections computed via {\tt AlpGen}, 
  with the parton distribution set CTEQ5L, and the  factorization  and renormalization  scale at the common value $\mu=m_H$.
  We assume the setup defined by Eqs. (\ref{cuts}) and (\ref{masses}), implying a cut 
  $p_T^{\gamma,j} >30$ GeV on the photon and (if present)  final-jets  transverse momenta. 
  In Table~\ref{tab:total}, we alternate columns referring to cross sections for main Higgs production channels (with no final photon), named $(H)_{\sqrt S}$, with the corresponding ones where an extra photon is required, named $(H\gamma)_{\sqrt S}$. 
  
  Note that the first process considered (first row, named $gg, qg ,\bar q q$), corresponding to the original  gluon-fusion Higgs  production,  includes in its {\it photon-less} $(H)_{\sqrt S}$ component  not only $gg\to H$, but also the Higgs+jet channel  proceeding at LO via the 
  $gg, qg (\bar q g),\bar q q \to H g, q(\bar q), g$ scattering, mediated by an effective 
  $ggH$ vertex. This is to match the corresponding $H\gamma$ top-loop component, which 
  requires at the lowest order an extra final parton in the processes 
  $gg, qg (\bar q g),\bar q q \to H\gamma\, g, q(\bar q), g$ 
  (as discussed in  Section 2.1)\footnote{For instance, at 14(100) TeV, the LO $gg, qg ,\bar q q$ component of $(H)_{\rm14(100)TeV}$ 
  [that is 30.8(745.)pb]  is made up of 19.7(415.)pb, coming from  the $gg\to H$ LO cross section,
   plus 7.9(274.)pb, arising from the  $gg \to H g$   LO cross section,
 plus 3.1(56.)pb,  from   $ qg (\bar q g) \to H q (\bar q)$, 
 with  negligible $q\bar q\to H g$ contributions.}.
 Note also that, the corresponding $(H\gamma)_{\sqrt S}$ component gets only a minor contribution from the pentagon $gg\to H\gamma g$ process (see  again Section 2.1).
  
  By looking at cross sections in Table \ref{tab:total}, it gets particularly clear that the requirement of an extra high-$p_T$ photon suppresses the original Higgs production rates
  by an amount
   that is widely dependent on the process.
  Top-loop production turns out to drop by a factor $10^{-4}$ at all c.m. energies considered, and is the most suppressed process. Slightly less suppressed 
  (by a factor $\sim 2.4 \cdot 10^{-4}$) is the $b\bar b\to H$ rate.
  On the contrary, both VBF and single-top  production loose just a factor $10^{-2}$ when adding a photon, and present the least decreased rates. The rates for all other channels drop by a few $10^{-3}$, with a suppression factor increasing going from  $c\bar c\to H$ ($10^{-3}$), up to 
   $WH,ZH$ ($\sim 1.3\cdot 10^{-3}-2 \cdot 10^{-3}$), and $t\bar t H$ ($\sim 4 \cdot 10^{-3}$).
   
   At the LHC, the most abundant $H\gamma$ production arises from VBF  (22 fb), with one-order-of-magnitude lower contributions from $VH$ (3.2 fb), top-loop production (3.1 fb), and $t\bar t H$
   (2.6 fb).
   At larger c.m. energies, VBF is still by far dominant, but the relative weight  of top-loop
   and direct top production increases considerably.
   At $\sqrt S\simeq 100$ TeV,  VBF is about 0.31 pb (\ie, more than 50\% of the total), $t\bar t H$ is about 0.16 pb, and all the  remaining $H\gamma$ channels sum up to about 0.12 pb.
  
\begin{figure}
\begin{center}
\vskip -0.5cm
\hskip -0.4cm
\includegraphics[width=0.5\textwidth]{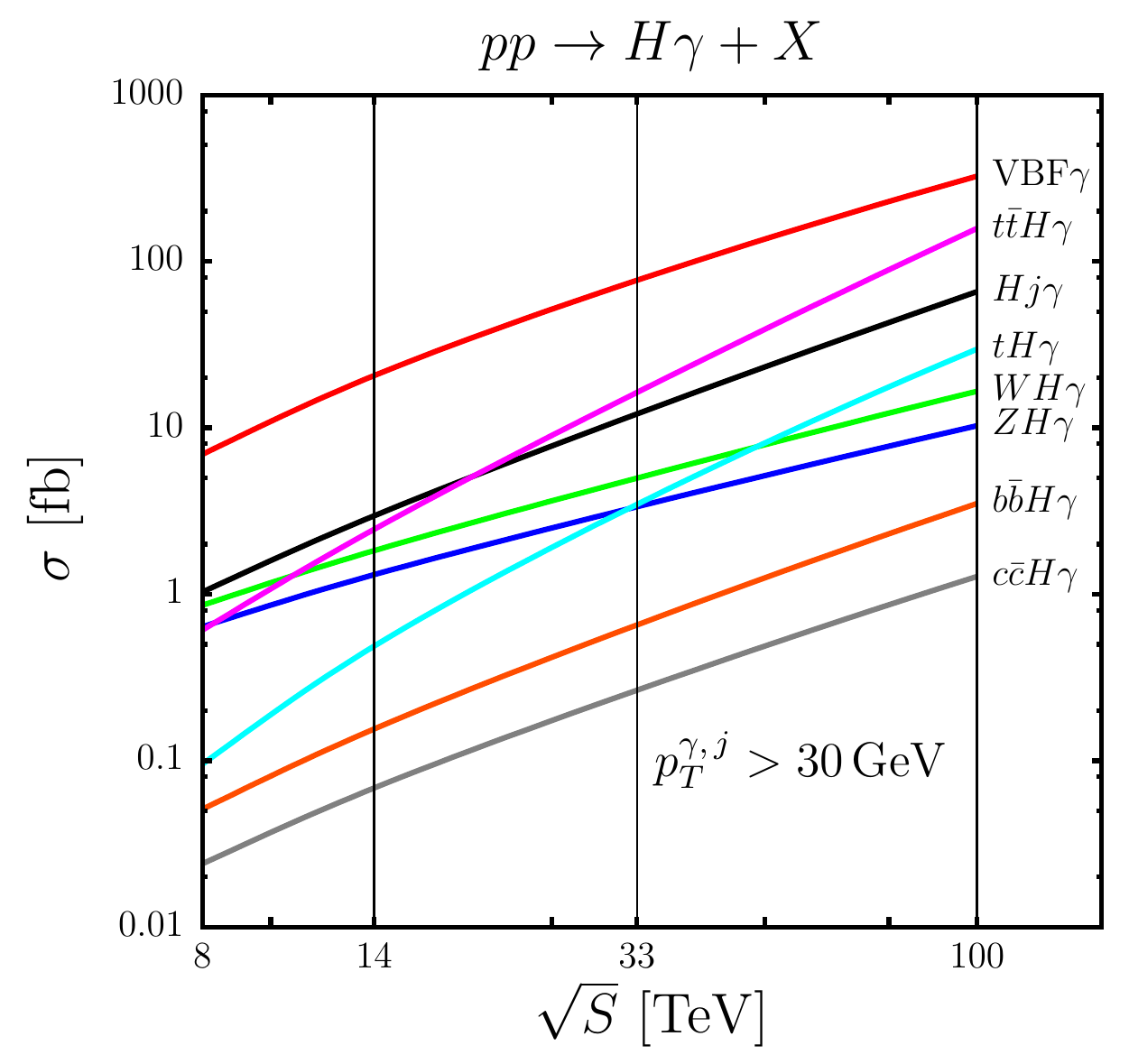}
\vskip -0.4cm
\caption{\small Cross sections for  $pp\to H  \gamma + X$ with same kinematical cuts as in Tables~\ref{tab:total} and \ref{tab:total2}.}
\label{xsecHG}
\end{center}
\end{figure}

\begin{table}
\begin{center}
\begin{tabular}{l|c|c|c|c|c|c|}
$\sigma_{(p_T^{\gamma,j} >30 {\rm GeV})}$
 & ($H)_{\rm 8TeV}$ &  ($H\gamma)_{\rm 8TeV}$ 
 & ($H)_{\rm 13TeV}$ &  ($H\gamma)_{\rm 13TeV}$
  \\ \hline

 $gg,gq,q \bar q $ & 10.3 pb & \; 1.03 fb & \; 26.8 pb &\; \; 2.68 fb 
 \\ 
 
 VBF & 0.844 &  6.93  \;     & 2.08 &  18.8 \;\;
 \\
 
 $WH$ & 0.552 &  0.858       & 1.07 &  1.70
  \\

$ZH$ & 0.291 & 0.637          & 0.567 & 1.23
 \\

$t\bar tH$ & 0.137 & 0.608       &  0.487  &  2.13  
 \\ 

$tH+\bar tH$ &  0.012 & 0.095       &  0.046  &   0.431
 \\ 

$b\bar b\to H$ & 0.232 & 0.051          &  0.586  &   0.140
 \\ 

$c\bar c\to H$ & 0.026 & 0.024          &   0.061  &   0.062
 \\  \hline
\end{tabular}
\end{center}
\caption{\small 
Same as in Table \ref{tab:total} at $\sqrt S= $8 TeV and 13 TeV.
}
\label{tab:total2}
\end{table}
\begin{table}
\vskip 0.5cm
\begin{center}
\begin{tabular}{l|c|c|c|c|c|c|}

 $\sigma_{(p_T^{\gamma,j} >20 {\rm GeV})}$ & ($H)_{\rm14TeV}$ &  ($H\gamma)_{\rm14TeV}$ 
  \\ \hline

 $gg,gq,q \bar q $ & \;\;35.7 pb & \;\;\;\; 4.61 fb  \\ 
 
 VBF & 3.02 &  38.8 \;\;      \\
 
 $WH$ & 1.17 &  2.85      \\

$ZH$ & 0.625 & 2.01      \\

$t\bar tH$ & 0.585 & 3.32      \\ 

$tH+\bar tH$ & 0.061 & 0.842        \\ 

$b\bar b\to H$ & 0.670 & 0.308      \\ 

$c\bar c\to H$ & 0.069 & 0.135         \\  \hline
\end{tabular}
\end{center}
\caption{\small 
Same as in Table \ref{tab:total} at $\sqrt S= $14 TeV,  and 
for $p_T^{\gamma,j} >20$ GeV.
}
\label{tab:total3}
\end{table}
 In Table \ref{tab:total2}, we present the corresponding rates at $\sqrt S\simeq 8$ and 13 GeV, with  same conventions as in Table~\ref{tab:total}. The $(H\gamma)_{\sqrt S}$
 cross sections, defined as in Tables~\ref{tab:total} and \ref{tab:total2}, for all processes versus $\sqrt{S}$ are also plotted in Figure~\ref{xsecHG}, which clearly shows 
 the new hierarchy of different Higgs production channels.

We stress that all the rates (and corresponding hierarchies) presented in 
Tables~\ref{tab:total} and \ref{tab:total2} somewhat depend on the kinematical selection of the final state. On the one hand, all rates are affected by the choice of the photon $p_T$ cut  (in general not  in a universal way). On the other hand, the channels including  jets among the final products
are also sensitive to the jet selection. A different selection can hence affect in principle the relative weight of channels. 

In Table \ref{tab:total3}, we present results at 14 TeV, when one relaxes the $p_T^{\gamma,j} >30$ GeV cuts in Eq.~(\ref{cuts}) down to the less strict selection $p_T^{\gamma,j} >20$ GeV, the latter being also quite realistic at the LHC energies.
The most affected channels are VBF and the heavy-quark fusion channels.
The former is quite dependent on  both  $p_T^{\gamma}$ and $p_T^{j}$ 
cuts, and as a consequence doubles its rate,  the latter are, among the processes considered, the most sensitive to the $p_T^{\gamma}$ cut.
    
\begin{figure}
\begin{center}
\hskip 0.1cm
\includegraphics[width=0.46\textwidth]{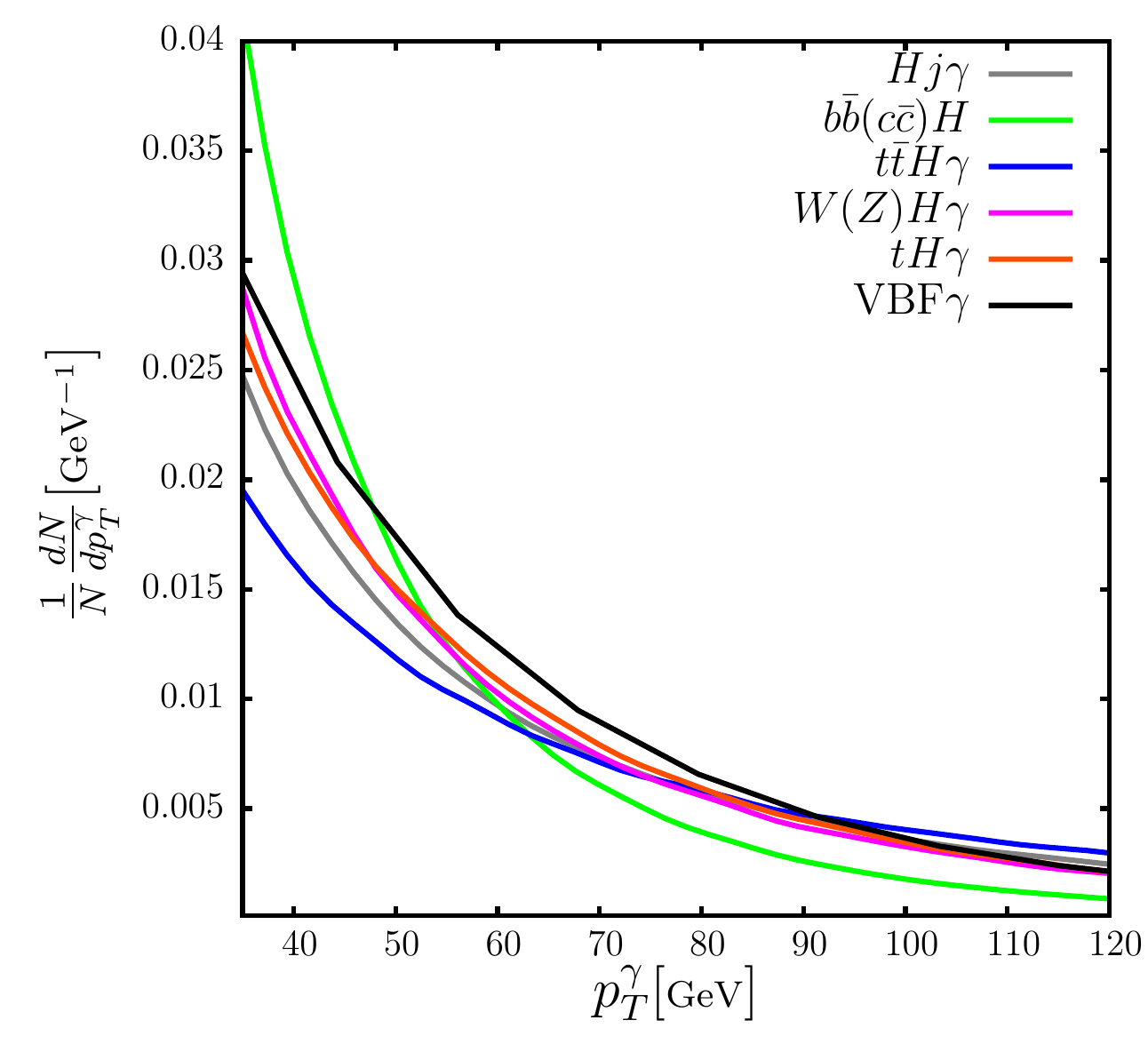}
\includegraphics[width=0.46\textwidth]{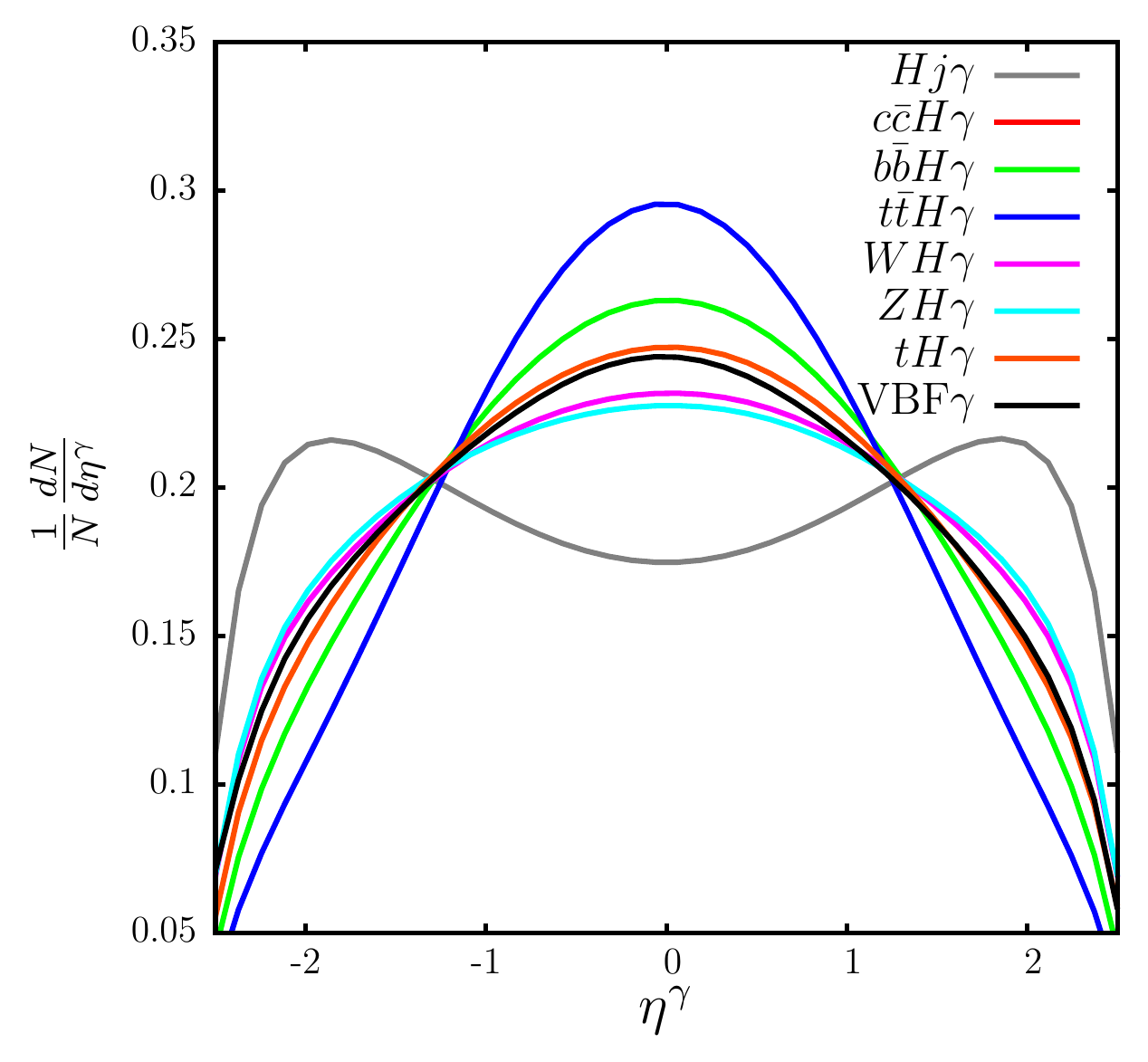}
\vskip -0.5cm
\caption{\small Photon transverse momentum and pseudorapidity distributions at $\sqrt S=$14 TeV. Conventions are detailed in the text.}
\label{fig_10}
\end{center}
\end{figure}
\begin{figure}
\begin{center}
\vskip -0.6cm
\hskip 0.1cm
\includegraphics[width=0.46\textwidth]{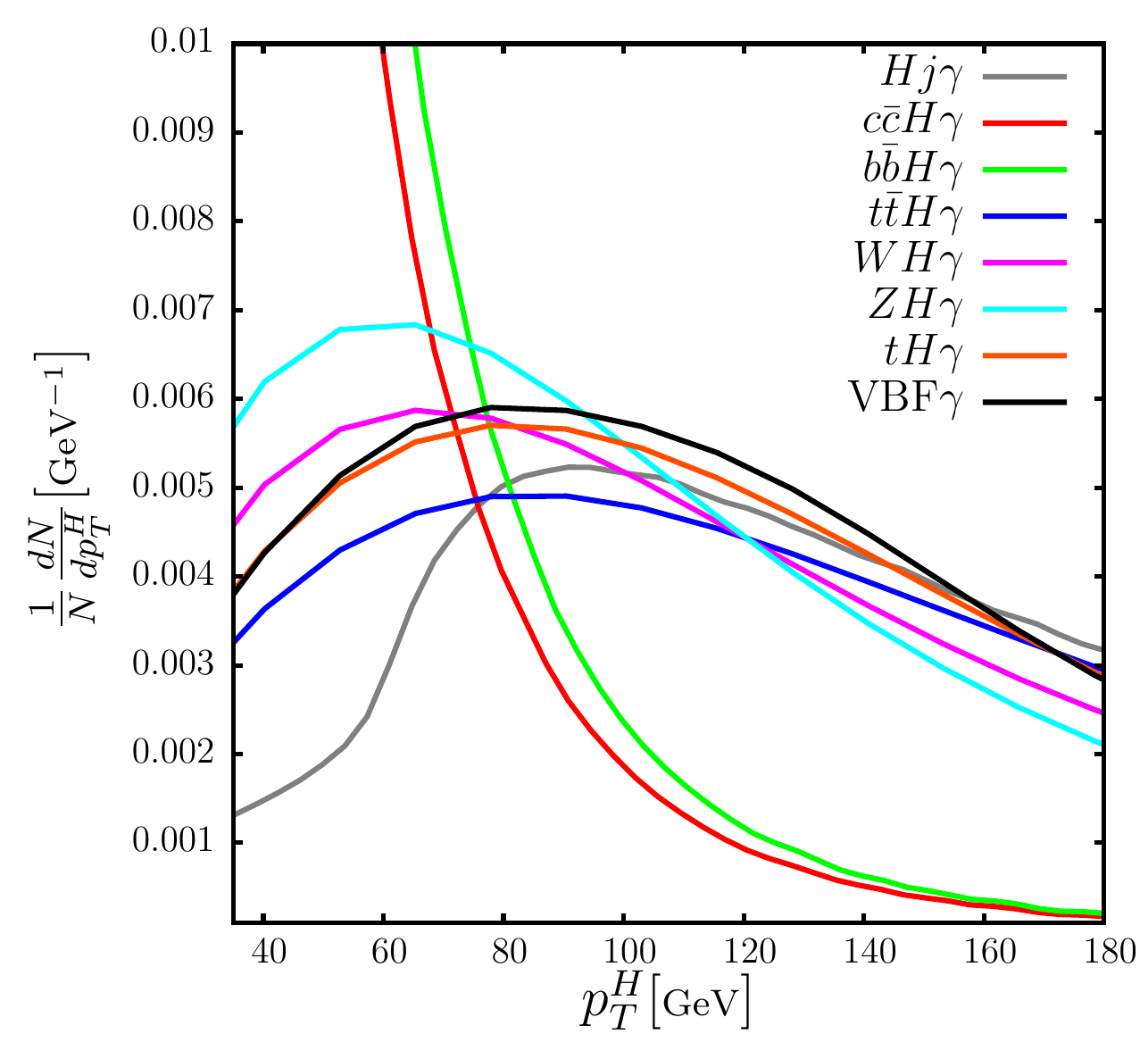}
\includegraphics[width=0.46\textwidth]{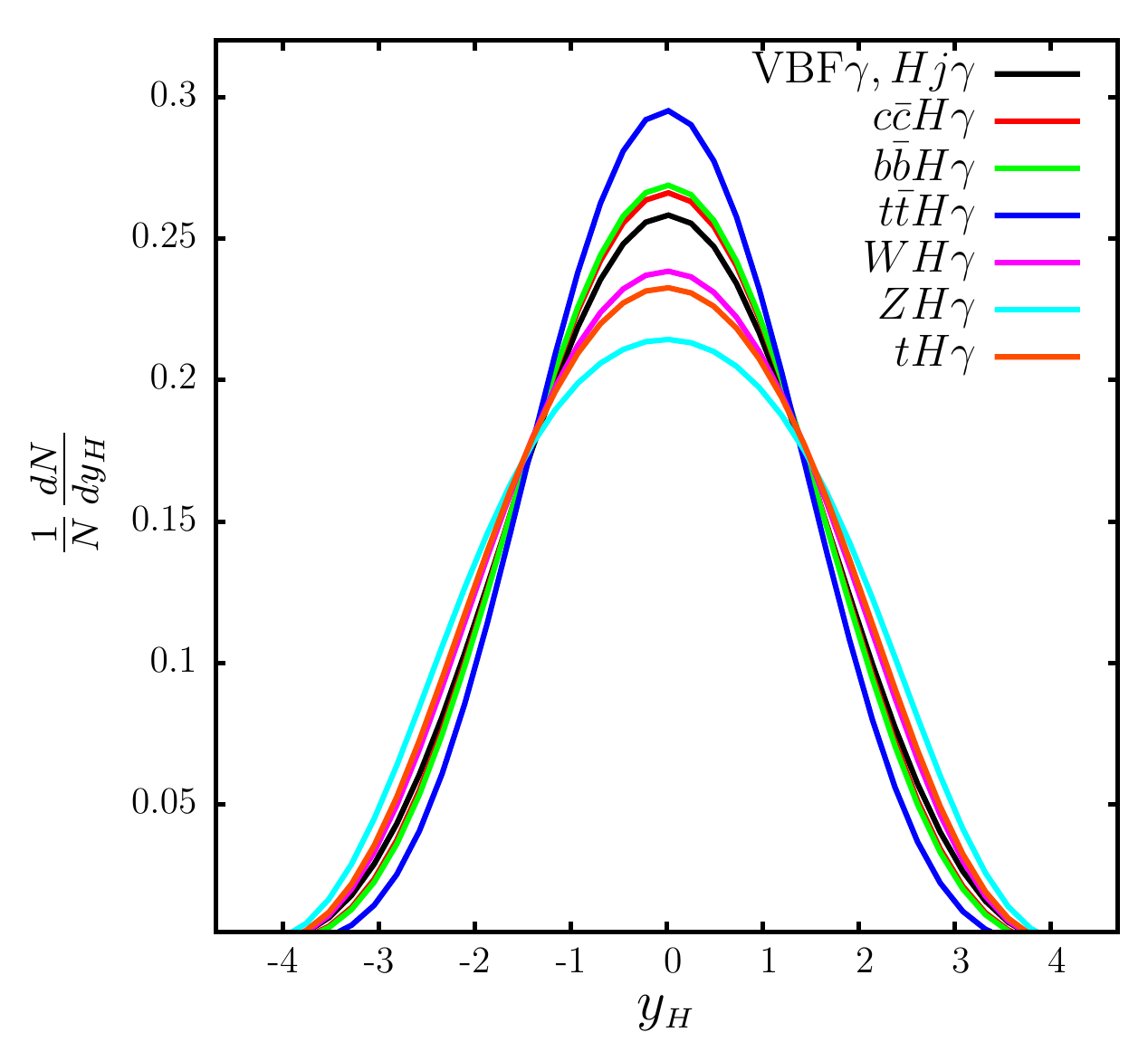}
\vskip -0.5cm
\caption{\small Higgs transverse momentum and rapidity distributions at $\sqrt S=$14 TeV. Conventions are detailed in the text.}
\label{fig_11}
\end{center}
\end{figure}
\begin{figure}
\begin{center}
\vskip -0.6cm
\hskip 0.1cm
\includegraphics[width=0.46\textwidth]{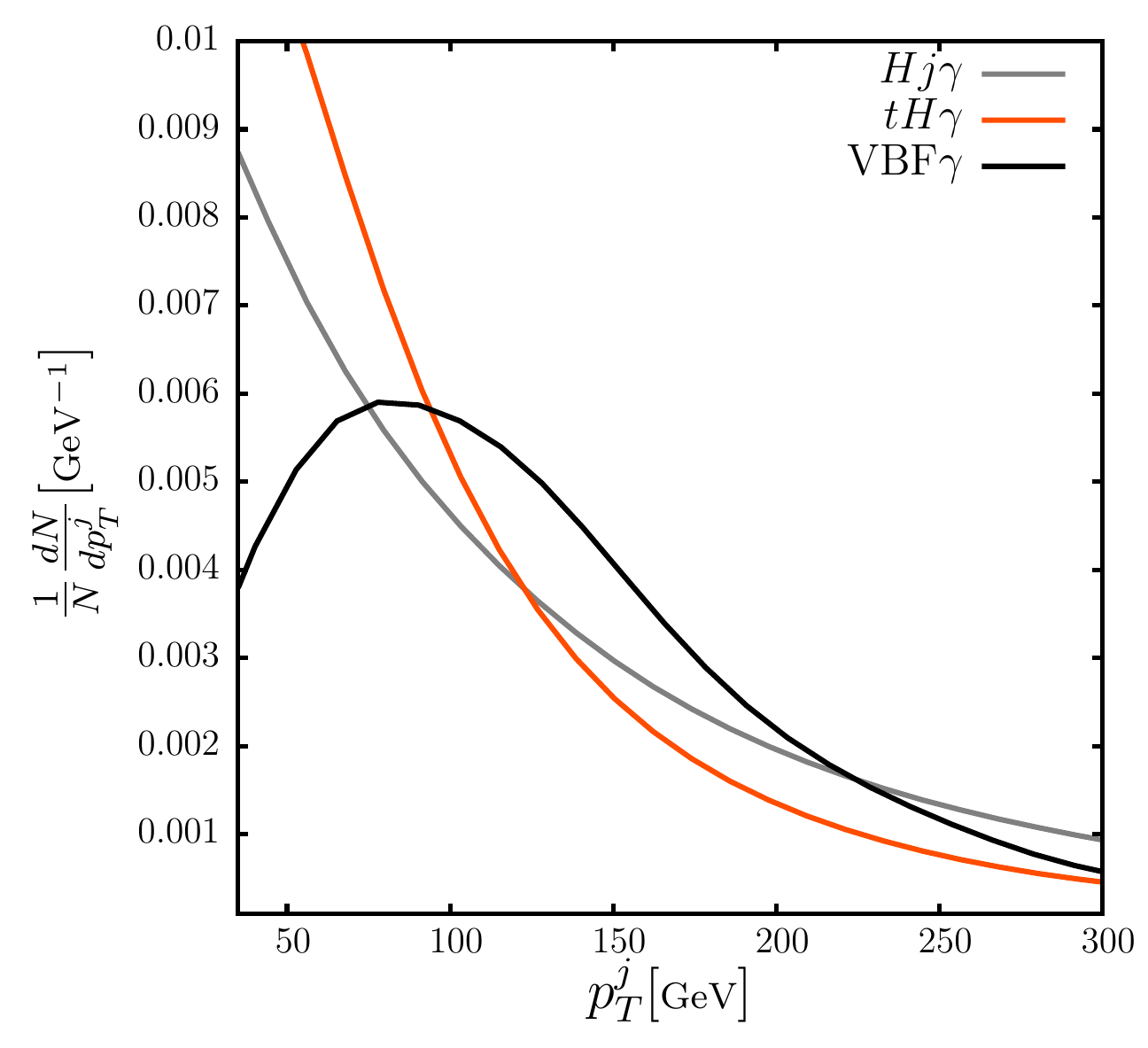}
\includegraphics[width=0.46\textwidth]{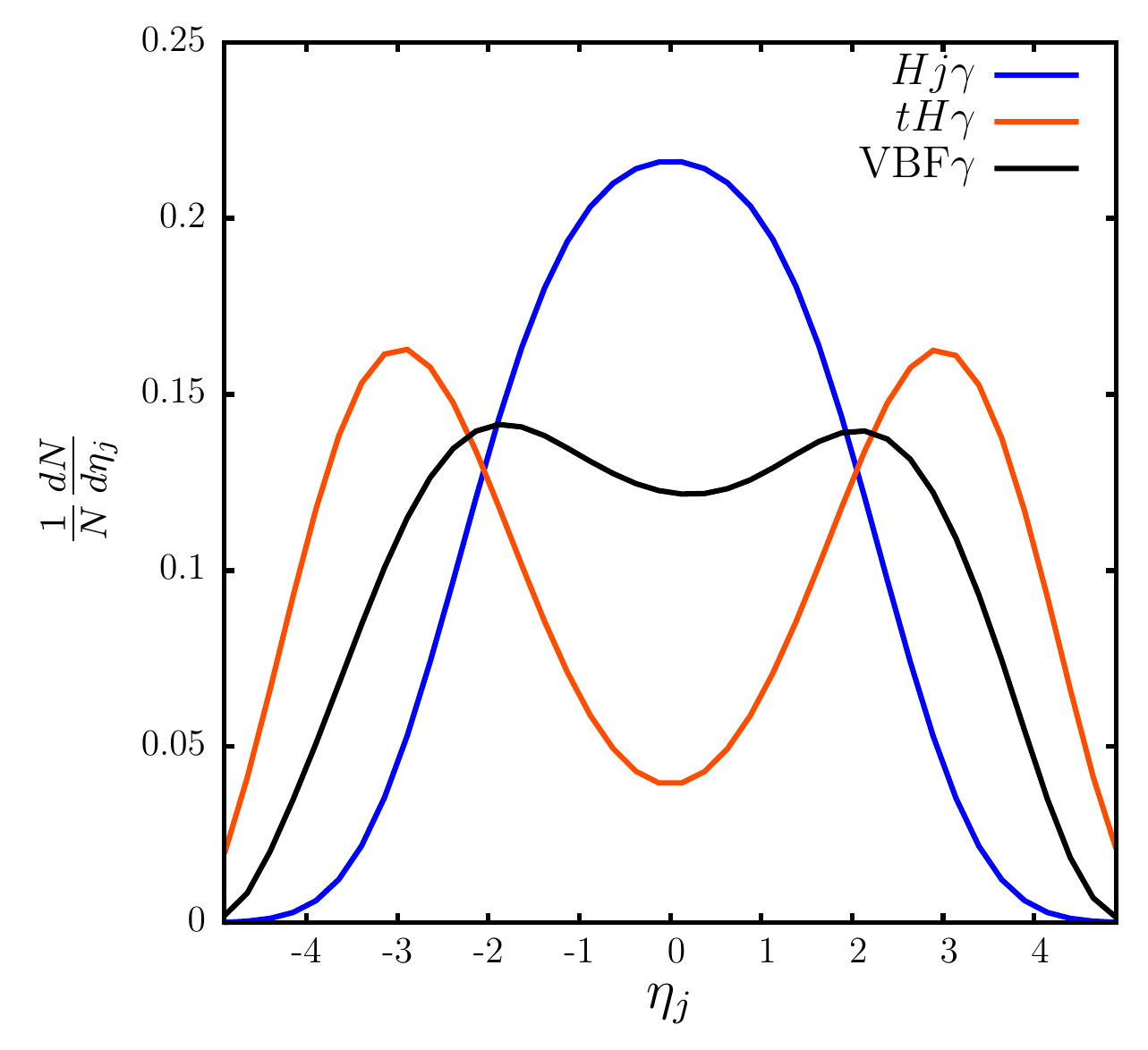}
\vskip -0.5cm
\caption{\small Jet transverse momentum and rapidity distributions at $\sqrt S=$14 TeV.
Conventions are detailed in the text.}
\label{fig_12}
\end{center}
\end{figure}
Indeed, the  impact of a change in the kinematical selection can be guessed by looking at the various kinematical 
distributions for the different  $H\gamma$ channels, which are shown in 
Figures~\ref{fig_10},
\ref{fig_11}, and \ref{fig_12}, for $\sqrt S=14$ TeV. All distributions are normalized 
to  unity, after applying the kinematical cuts in Eq.~(\ref{cuts}). 
Distributions detailed by the   lines named ``$Hj\gamma$" refer to the $gg,gq,q \bar q $ channels mediated by the effective $ggH$ vertex.
One can see that the rate  dependence on the photon, Higgs, and jet momenta of various processes can partly alter their relative weight when changing the kinematical selection.
In Figure~\ref{fig_10}, the heavy-quark fusion $b\bar b, c\bar c\to H\gamma $ channels  present the steepest 
dependence on $p_T^\gamma$, while $H\gamma \,t\bar t$ shows the mildest
dependence among the channels considered. The $p_T^H$ dependence in 
Figure~\ref{fig_11} is somewhat more structured. All the processes but $b\bar b, c\bar c\to H\gamma $  show the  maximum of $p_T^H$ distributions at a quite large 
 $p_T^H$ value. The VBF$\gamma$ channel (where by VBF$\gamma$
we name the $H\gamma$ production via VBF) shows the typical $p_T^H\sim M_W$
maximum, which is also present in the basic VBF Higgs production. On the other hand, the $Hj\gamma$ 
channel, mostly arising from $qg\to H\gamma q$, has an even higher average $p_T^H$, since 
in this case the photon tends to be collinear with the initial/final quark, and the Higgs boson $p_T$  has to balance
the $p_T$ of the $j\gamma$ system, each component of which is required to have   $p_T>30$~GeV.
In Figure~\ref{fig_12}, we detail the jet distributions for the few processes where at least one jet is present in the final state.

\section{Conclusions}
We have made a general analysis of the processes giving rise to final states
containing a Higgs boson and a high-$p_T$ photon in proton collisions at different c.m. energies,  relevant at the LHC and future colliders.
We showed that the request of an extra photon in Higgs production widely affects the normal hierarchy in the main Higgs production processes.
In particular, most of the $H\gamma$ signal derives in general from the VBF production. 

At the LHC, for  $p_T^{\gamma,j} >20$ GeV, VBF accounts for more than 70\% 
of the $H\gamma$ final states  in a LO analysis. The second most important process
is the one mediated by the top-loop effective $ggH$ coupling which is responsible for about 9\% of the production, although  contributing only slightly more than  the $t\bar tH $
direct top production, and a bit less than the total $WH/ZH$ associated production. 
At larger c.m. energies (in particular at  $\sqrt S\gsim 33$ TeV), $t\bar tH $ gets the upper hand, and becomes the second (following VBF) most relevant process.

As a result,  asking for an extra photon in  Higgs production reverses the order of importance of the gluon-fusion and VBF mechanisms.
Remarkably, the emission of the photon highly suppress the $ZZ$ fusion component with respect to the $WW$ one in the VBF channel \cite{Gabrielli:2007wf}.
Hence, the requirement of an extra photon in inclusive  Higgs production 
naturally selects samples with good purity of the $WW$ VBF component,
with a rate suppression factor of the order 10$^{-2}$ with respect to the main VBF Higgs channel.

In the present study, we have redone the computation of the $gg\to H \gamma g$
pentagon amplitude, confirming results recently appeared in the literature. 
On the other hand, we corrected the relative weight previously assigned to this channel in the   production of $H\gamma j $ final states, pointing out the dominant role  of   processes mediated by the $ggH$ effective coupling. 
We also computed for the first time the associated Higgs and photon production in processes involving 
direct production of top pairs and single top.

Of course, the present analysis would be made more robust by a general NLO treatment of all processes. We anyway expect this refinement to keep the general features of the present discussion unchanged.

Accurate predictions  for the associated production of a Higgs boson and a photon at the LHC will  be crucial not only to test $H\gamma$ interactions, but also in
   probing new physics effects in the associated production of new scalar particles and  photons  \cite{Li:2012zzs},
as well as  in  searching for  resonant three-photon final states \cite{Toro:2012sv}, \cite{Aad:2015bua}. For instance, in case of the production of a scalar $\phi$ with features departing from the Higgs-boson ones, the present cross-section hierarchy could be widely affected. Non-vanishing terms like $gg\phi\gamma$  in the effective-Lagrangian interactions  (violating $C$ parity in the SM  case) might resurrect the role of the gluon fusion process as a dominant  production channel for the scalar-photon final state, and change the overall picture.

\acknowledgments
This research was supported in part by the European Commission through contracts 
ERC-2011-AdG No 291377 (LHCtheory) and PITN-GA-2012-316704 (HIGGSTOOLS) and 
by the Italian Ministry of University and Research under the PRIN project 2010YJ2NYW.
R.P. also thanks the project FPA2013-47836-C3-1-P. 
E.G, F.P., and R.P. acknowledge the CERN TH-Unit for its hospitality and partial support 
during the preparation of this work. The authors thank the Galileo Galilei Institute for Theoretical 
Physics for hospitality 
and INFN for partial support while part of this work was carried out. F.P. would like to thank 
the Mainz Institute for 
Theoretical Physics (MITP) for hospitality and support while part of this work was carried out.

\appendix

\section{Asymptotic behavior of the top pentagon amplitude for $gg\to H \gamma g$.}  
The $g(p_1) g(p_2) g(p_3) \gamma(p_4) \to H$ amplitude that is relevant for the process in Figure~\ref{fig_1} is proportional to
\begin{eqnarray}
&&A^{\mu_1\mu_2\mu_3\mu_4}(p_1,p_2,p_3,p_4) =   
\frac{m_t}{i \pi^2}\sum_{\sigma \in S_4(\left\{1,2,3,4\right\}) }
\int d^4q\, {\rm Tr} \left\{
\frac{1}{\rlap/ q_{\sigma(4)}-m_t}
 \gamma^{\mu_{\sigma(3)}}
 \right. \nonumber \\
&&\hspace{20pt}\times \left.
\frac{1}{\rlap/ q_{\sigma(3)}-m_t}
 \gamma^{\mu_{\sigma(2)}}
\frac{1}{\rlap/ q_{\sigma(2)}-m_t}
 \gamma^{\mu_{\sigma(1)}}
\frac{1}{\rlap/ q_{\sigma(1)}-m_t}
 \gamma^{\mu_{\sigma(4)}} 
\frac{1}{\rlap/ q-m_t}
\right\},
\end{eqnarray}
with
\begin{equation}
q_{\sigma(j)}= q + p_{\sigma(4)}+ \sum_{i=1}^{j-1}p_{\sigma(i)}.
\end{equation}
For $m_t\gg m_H,\sqrt s$, the pentagon amplitude is well approximated by the
following simple expression
\begin{eqnarray}
&&A_{m_t\to \infty}^{\mu_1\mu_2\mu_3\mu_4}(p_1,p_2,p_3,p_4) \sim -\frac{1}{m_t^4}
\sum_{\substack{j=2 \\ 1 < k \ne j \\  k < l}}^{4}
\bigg\{
 \frac{32}{9} T_2^{\mu_1\mu_j}(p_1,p_j)T_2^{\mu_k\mu_l}(p_k,p_l) \nonumber \\
&& \hspace{20pt}- \frac{112}{45}T_4^{\mu_1\mu_j\mu_k\mu_l}(p_1,p_j,p_k,p_l)
\bigg\},
\end{eqnarray}
with
\begin{eqnarray}
T_2^{\mu_i\mu_j}(p_i,p_j)= g^{\mu_i\mu_j}(p_i \cdot p_j)-p_i^{\mu_j}p_j^{\mu_i}
\end{eqnarray}
and
\begin{eqnarray}
T_4^{\mu_1\mu_j\mu_k\mu_l}(p_1,p_j,p_k,p_l) &=& 
  p_1^{\mu_k} p_j^{\mu_l} p_k^{\mu_j} p_l^{\mu_1}
+ p_1^{\mu_l} p_j^{\mu_k} p_k^{\mu_1} p_l^{\mu_j} \nonumber \\
&&\hspace{-77pt}+\ g^{\mu_1\mu_j}g^{\mu_k\mu_l} \big[(p_1 \cdot p_k)(p_j \cdot p_l)
                            +(p_1 \cdot p_l)(p_j \cdot p_k)\big] \nonumber \\
&&\hspace{-77pt}+ \ g^{\mu_1\mu_j}
\big\{
           (p_1^{\mu_k}  p_j^{\mu_l}+p_1^{\mu_l} p_j^{\mu_k}) (p_k\cdot p_l)
               - p_k^{\mu_l} [p_1^{\mu_k}(p_j\cdot p_l)+p_j^{\mu_k}(p_1\cdot p_l)]
\nonumber \\
&&\hspace{10pt}- p_l^{\mu_k} [p_1^{\mu_l}(p_j\cdot p_k)+p_j^{\mu_l}(p_1\cdot p_k)]
\big\} \nonumber \\
&&\hspace{-77pt}+\ g^{\mu_k\mu_l}
\big\{
           (p_k^{\mu_1}  p_l^{\mu_j}+p_k^{\mu_j} p_l^{\mu_1}) (p_1\cdot p_j)
               - p_1^{\mu_j} [p_k^{\mu_1}(p_j\cdot p_l)+p_l^{\mu_1}(p_j\cdot p_k)]
\nonumber \\
&&\hspace{10pt}- p_j^{\mu_1} [p_k^{\mu_j}(p_1\cdot p_l)+p_l^{\mu_j}(p_1\cdot p_k)]
\big\}. 
\end{eqnarray}



\begin{thebibliography}{99}
%
\bibitem{Aad:2012tfa}
  G.~Aad {\it et al.}  [ATLAS Collaboration],
  Phys.\ Lett.\ B {\bf 716}, 1 (2012)
  [arXiv:1207.7214 [hep-ex]];
  S.~Chatrchyan {\it et al.}  [CMS Collaboration],
  Phys.\ Lett.\ B {\bf 716}, 30 (2012)
  [arXiv:1207.7235 [hep-ex]].


\bibitem{Englert:1964et}
  F.~Englert and R.~Brout,
  Phys.\ Rev.\ Lett.\  {\bf 13}, 321 (1964);
  P.~W.~Higgs,
  Phys.\ Lett.\  {\bf 12}, 132 (1964);
  P.~W.~Higgs,
  Phys.\ Rev.\ Lett.\  {\bf 13}, 508 (1964);
  G.~S.~Guralnik, C.~R.~Hagen and T.~W.~B.~Kibble,
  Phys.\ Rev.\ Lett.\  {\bf 13}, 585 (1964).
 
    \bibitem{H-xsect}
     LHC Higgs Cross Section Working Group, 
       {\sl Handbook of LHC Higgs Cross Sections: 1. Inclusive Observables},     CERN-2011-002 (CERN, Geneva, 2011), [arXiv:1101.0593 [hep-ph]]; 
     {\sl 
     2. Differential Distributions}, 
     CERN-2012-002 (CERN, Geneva, 2012), [arXiv:1201.3084 [hep-ph]],  
     S.~Dittmaier, C.~Mariotti, G.~Passarino, and R.~Tanaka (Eds.), 
 {\sl  3. Higgs Properties},
 CERN-2013-004 (CERN, Geneva, 2013), [arXiv:1307.1347 [hep-ph]],
   S.~Heinemeyer, C.~Mariotti, G.~Passarino, and R.~Tanaka (Eds.). 
     
%
\bibitem{Agrawal:2014tqa}
  P.~Agrawal and A.~Shivaji,
  Phys.\ Lett.\ B {\bf 741} (2015) 111
  [arXiv:1409.8059 [hep-ph]].
  
    
\bibitem{Gabrielli:2007wf}
  E.~Gabrielli, F.~Maltoni, B.~Mele, M.~Moretti, F.~Piccinini and R.~Pittau,
  Nucl.\ Phys.\ B {\bf 781} (2007) 64
  [hep-ph/0702119 [HEP-PH]].

\bibitem{Arnold:2010dx}
  K.~Arnold, T.~Figy, B.~Jager and D.~Zeppenfeld,
  JHEP {\bf 1008} (2010) 088
  [arXiv:1006.4237 [hep-ph]].


\bibitem{Mao:2013dxa}
  S.~Mao, W.~Neng, L.~Gang, M.~Wen-Gan, Z.~Ren-You, G.~Lei, Z.~Ya-Jin and G.~Jian-You,
  Phys.\  Rev.\  D {\bf 88:076002} (2013)
  [arXiv:1310.0946 [hep-ph]].

\bibitem{Shou-Jian:2015sta}
  X.~Shou-Jian, M.~Wen-Gan, G.~Lei, Z.~Ren-You, C.~Chong and S.~Mao,
  J.\ Phys.\ G {\bf 42} (2015) 6,  065006
  [arXiv:1505.03226 [hep-ph]].

\bibitem{Abbasabadi:1997zr}
  A.~Abbasabadi, D.~Bowser-Chao, D.~A.~Dicus and W.~W.~Repko,
  Phys.\ Rev.\ D {\bf 58} (1998) 057301
  [hep-ph/9706335].


\bibitem{Mangano:2002ea}
  M.~L.~Mangano, M.~Moretti, F.~Piccinini, R.~Pittau and A.~D.~Polosa,
  JHEP {\bf 0307} (2003) 001
  [hep-ph/0206293].

\bibitem{Lai:1999wy}
  H.~L.~Lai {\it et al.} [CTEQ Collaboration],
  Eur.\ Phys.\ J.\ C {\bf 12} (2000) 375
  doi:10.1007/s100529900196
  [hep-ph/9903282].
  
\bibitem{fcc}
ÒFuture Circular Collider Study Kickoff Meeting, Geneva, 12-15 February 2014Ó.
https://indico.cern.ch/event/282344/
.

\bibitem{Arkani-Hamed:2015vfh}
  N.~Arkani-Hamed, T.~Han, M.~Mangano and L.~T.~Wang,
  arXiv:1511.06495 [hep-ph].
  
\bibitem{Baglio:2015wcg}
  J.~Baglio, A.~Djouadi and J.~Quevillon,
  arXiv:1511.07853 [hep-ph].

\bibitem{Andreazza:2015bja}
  A.~Andreazza {\it et al.},
  Frascati Phys.\ Ser.\  {\bf 60} (2015) 1.

\bibitem{Ossola:2006us}
  G.~Ossola, C.~G.~Papadopoulos and R.~Pittau,
  Nucl.\ Phys.\ B {\bf 763} (2007) 147
  [hep-ph/0609007].

\bibitem{Ossola:2007ax}
  G.~Ossola, C.~G.~Papadopoulos and R.~Pittau,
  JHEP {\bf 0803} (2008) 042
  [arXiv:0711.3596 [hep-ph]].
  
\bibitem{vanHameren:2010cp}
  A.~van Hameren,
  Comput.\ Phys.\ Commun.\  {\bf 182} (2011) 2427
  [arXiv:1007.4716 [hep-ph]].


\bibitem{Ossola:2008xq}
  G.~Ossola, C.~G.~Papadopoulos and R.~Pittau,
  JHEP {\bf 0805} (2008) 004
  [arXiv:0802.1876 [hep-ph]].


\bibitem{Kleiss:1985yh}
  R.~Kleiss and W.~J.~Stirling,
  Nucl.\ Phys.\ B {\bf 262} (1985) 235.
  
\bibitem{Torrielli:2014rqa}
  P.~Torrielli,
  arXiv:1407.1623 [hep-ph].
  
\bibitem{Maltoni:2001hu}
  T.~M.~P.~Tait and C.-P.~Yuan,
  Phys.\ Rev.\ D {\bf 63} (2000) 014018
  [hep-ph/0007298];
  F.~Maltoni, K.~Paul, T.~Stelzer and S.~Willenbrock,
  Phys.\ Rev.\ D {\bf 64} (2001) 094023
  [hep-ph/0106293];
  V.~Barger, M.~McCaskey and G.~Shaughnessy,
  Phys.\ Rev.\ D {\bf 81} (2010) 034020
  [arXiv:0911.1556 [hep-ph]].
  
\bibitem{Biswas:2012bd}
  S.~Biswas, E.~Gabrielli and B.~Mele,
  JHEP {\bf 1301} (2013) 088
  [arXiv:1211.0499 [hep-ph]]; 
  M.~Farina, C.~Grojean, F.~Maltoni, E.~Salvioni and A.~Thamm,
  JHEP {\bf 1305} (2013) 022
  [arXiv:1211.3736 [hep-ph]];
  S.~Biswas, E.~Gabrielli, F.~Margaroli and B.~Mele,
  JHEP {\bf 1307} (2013) 073
  [arXiv:1304.1822 [hep-ph]].

\bibitem{Gabrielli:2007zp}
  E.~Gabrielli, B.~Mele and J.~Rathsman,
  Phys.\ Rev.\ D {\bf 77} (2008) 015007
  doi:10.1103/PhysRevD.77.015007
  [arXiv:0707.0797 [hep-ph]].



\bibitem{Li:2012zzs}
  X.~X.~Li, C.~X.~Yue, J.~X.~Chen and J.~N.~Dai,
  Chin.\ Phys.\ C {\bf 36} (2012) 485.

\bibitem{Toro:2012sv}
  N.~Toro and I.~Yavin,
  Phys.\ Rev.\ D {\bf 86} (2012) 055005
  doi:10.1103/PhysRevD.86.055005
  [arXiv:1202.6377 [hep-ph]].


\bibitem{Aad:2015bua}
  G.~Aad {\it et al.} [ATLAS Collaboration],
  arXiv:1509.05051 [hep-ex].
  
\end{thebibliography}
\end{document}